%% file: msaph.tex
\documentclass[usenatbib]{HAWC2015}
\usepackage{mathptmx}       
\usepackage{helvet}         
\usepackage{courier}        
\usepackage{type1cm}        
\usepackage[bottom]{footmisc}

\usepackage{natbib}
\usepackage{graphicx}        
\newcommand{\prd}{Physical Review D}   
\input{macos.tex}

\begin{document}

\title*{Gamma Ray Bursts in the HAWC Era}
%
\titlerunning{GRB in the HAWC era}
\author{Peter M\'esz\'aros$^{1,3}$, Katsuaki Asano$^{2}$, Kohta Murase$^{1}$, Derek Fox$^{1}$, He Gao$^{1}$, Nicholas Senno$^{1}$
}
\authorrunning{M\'esz\'aros et al.} 

\institute{
$^{1}$Dept. of Astronomy \& Astrophysics and Dept. of Physics, Center for Particle and
Gravitational Astrophysics, Pennsylvania State University, University Park, PA 16802, USA\\
$^{2}$Institute for Cosmic Ray Research, The University of Tokyo,
5-1-5 Kashiwanoha, Kashiwa, Chiba 277-8582, Japan\\
$^{3}$Date: May 18, 2015. Based on an invited talk at the HAWC inauguration conference, Puebla, Mexico, March 19, 2015.  To appear in Pubs. Academia de Ciencias de M\'exico, Eds. Alberto Carrami\~nana, Daniel Rosa, Stefan Westerhoff, Omar Tibolla\\
}
\maketitle
\vskip -3. cm  
\abstract{
Gamma-Ray Bursts are the most energetic explosions in the Universe, and
are among the most promising for detecting multiple non-electromagnetic
signals, including cosmic rays, high energy neutrinos and gravitational
waves.  The multi-GeV to TeV gamma-ray range of GRB could have significant
contributions from hadronic interactions, mixed with more conventional
leptonic contributions.  This energy range is important for probing the
source physics, including overall energetics, the shock parameters and
the Lorentz factor.  We discuss some of the latest observational and
theoretical developments in the field.
}
\vskip -0.5 cm
{\setlength{\unitlength}{1.cm}

\section{Introduction}
\label{sec:intro}

Gamma ray bursts have continued to astonish and puzzle us since the early 1970's
when they were first publicly discussed. After it became clear that they were
extragalactic, and hence had to be the most powerful and concentrated explosive
events in the Universe, which furthermore concentrate most of their output at 
MeV or higher energies, they have become a favored laboratory for exploring the 
most extreme environments and the highest energy processes in Nature. As such,
they are the targets for increasingly sophisticated observing instruments, the
latest of which is HAWC.

\section{The basic GRB paradigm}
\label{sec:basic}

In the course of the past 20-25 years of investigations, it has emerged that gamma-ray 
bursts (GRBs) can be classified in two main groups, ``long" GRBs (LGRBs) and ``short" GRBs 
(SGRBs) (e.g. \cite{Vedrenne+09book}). The former have spectral peaks on average below
0.5 MeV and $\gamma$-ray light curves whose duration nominally exceeds 2 s, while the latter 
have spectral peaks on average above 0.5 MeV and durations which mostly are less than 2 s. 
The stellar progenitors of LGRBs, which are generally located in star-forming galaxies,
appear to be massive stars ($M\simg 28\msun$), whose core collapses after they have
gone through their nuclear fuel burning cycle, while the progenitors of SGRBs are most 
likely compact binaries (double neutrons stars or neutron star-black hole binaries) whose 
merger is driven by gravitational wave emission (e.g. \cite{Woosley+06araa,Gehrels+09araa}).
See Fig. \ref{fig:longshort}.
\begin{figure}[ht]
  \centering
  \includegraphics[width=12 cm]{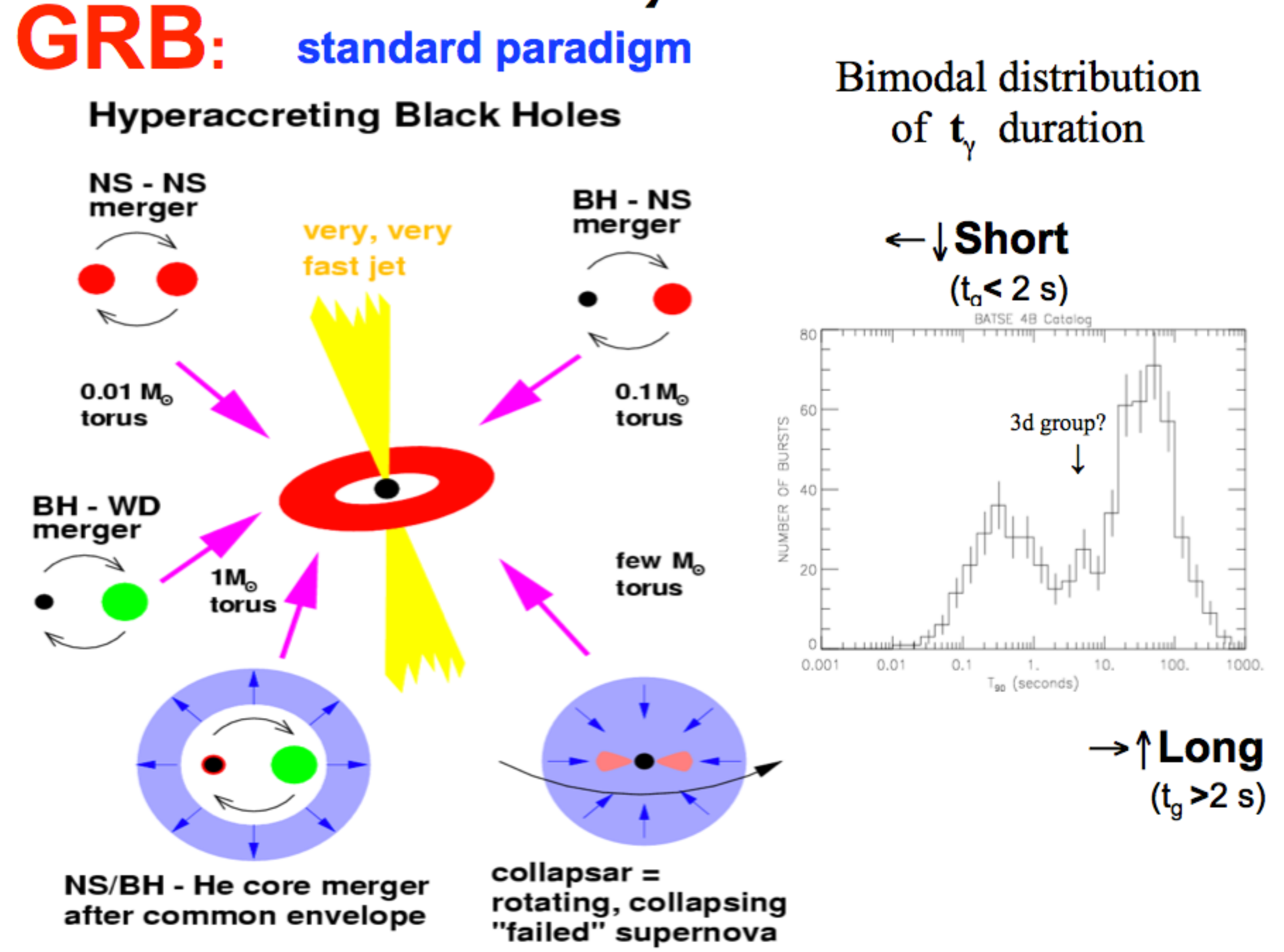}
  \caption{
Progenitors of long and short GRBs, after M. Ruffert and H.-Th. Jahnka (1996),
duration distribution from \cite{Kouveliotou+93}.
}
\label{fig:longshort}
\end{figure}

The mechanism by which these stellar cataclysmic events liberate such huge energies ($\sim
10^{54}\erg$) in such short times (down to milliseconds) is ultimately linked to the liberation
of the gravitational energy involved in  the compact stellar core or in the merging of
two degenerate stars (in both cases in a region of order a few Schwarzschild radii
corresponding to a few solar masses withing a dynamic time of order milliseconds),
and its conversion into thermal energy plus a final compact object, which ends up as
a black hole. This thermal energy, produced explosively, manifests itself initially
as a hot fireball (virial temperatures in the MeV range) consisting of photons, electrons,
positrons, nuclei and magnetic fields, as well as neutrinos and gravitational waves. The prompt 
neutrinos produced are predominantly thermal, in the 10-20 MeV range, similar to those expected 
in core collapse supernova, and amount to a few$\times 10^{53}\erg$ of energy, while the
gravitational wave emission is mainly in the $10^2-10^3\Hz$ range, amounting again to
a few$\times 10^{53}\erg$ for SGRBs, down to possibly much less for LGRBs, depending on
the variability of the total gravitational quadrupole moment during the merger or collapse.
Both these neutrinos and the gravitational waves escape in a matter of seconds or less -
the progenitor star is essentially transparent to them. 

The very high optical depth, however, traps the fireball photons, $e^\pm$, nucleons and 
magnetic fields, and this trapped thermal energy leads to an accelerating expansion of 
the fireball, converting thermal into bulk kinetic energy.  For an initial ratio of
energy to rest mass $\eta=E_0/M_0 c^2 \gg 1$ the ultimately resulting  bulk Lorentz factor 
eventually can become very large, $\Gamma \siml \eta$. The observation of GeV photons 
would imply thermal spectra below the pair threshold $\sim 0.5\MeV$, unless the expansion
Lorentz factor is very large $\Gamma\sim 10^2-10^3$ (e.g. \cite{Piran04grbrev,Meszaros06grbrev,
Gehrels+09araa}), which in turn requires very low baryon loads in the fireball. The observations, 
in fact, indicate that this relativistic expansion occurs along a collimated jet, rather than  
isotropically \citep{Gehrels+09araa}. 

If the photons escaped when the high optical depth had decreased to below unity one would
have expected both a low radiative efficiency (most energy is bulk kinetic) and a quasi-thermal
spectrum. Both issues, however, are avoided if the radiation arises in shocks, which are
expected to occur in the outflow, e.g. \cite{Meszaros06grbrev}. 
Shocks achieve the reconversion of kinetic energy into random particle energy, and Fermi 
acceleration in shocks leads to relativistic power law particle distributions which in
the magnetic fields  amplified in the shocks lead to synchrotron and inverse Compton non-thermal
spectra, when the shock occur beyond the scattering photosphere \citep{Rees+92fbal,Rees+94unsteady}. 
In addition, a quasi-thermal contribution, possibly with a non-thermal tail, can be expected 
from the photosphere, which however to be important requires internal dissipation. 
One expects internal shocks in the outflow when shells ejected with some variability timescale
$t_{v}$ with higher $\Gamma$ catch up with slower $\Gamma$ shells around a radius
$r_{is}\simeq 2 c t_{v}\Gamma^2 \sim 10^{12}t_{var,-3}\Gamma_2^2\cm$. An external (or termination) 
shock  is expected when the jet starts to be decelerated by a swept up external matter of density
$n_0$ (either ISM or a progenitor wind), at a radius $r_{es}\simeq (3E_0/4\pi n_0 m_pc^2)^{1/3}
\Gamma^{-2/3}\sim 3\times 10^{16}(E_{51}/n_0)^{1/3}\Gamma_2^{-3/2}\cm$, see
Fig \ref{fig:jetshock}. The internal shocks (which if magnetic fields are tangled could
be a magnetic dissipation region) occur closer in, and are highly variable and prominent 
at $\gamma$-ray energies \citep{Rees+94unsteady}. The external shocks occur further out,
producing, as the jet is increasingly decelerated, an increasingly softer and longer duration
afterglow spectrum, from X-rays through optical to radio \citep{Meszaros+97ag}.
A qualitatively similar spectral-temporal behavior is expected for both long and short bursts.
\begin{figure}[ht]
  \centering
  \includegraphics[width=12 cm]{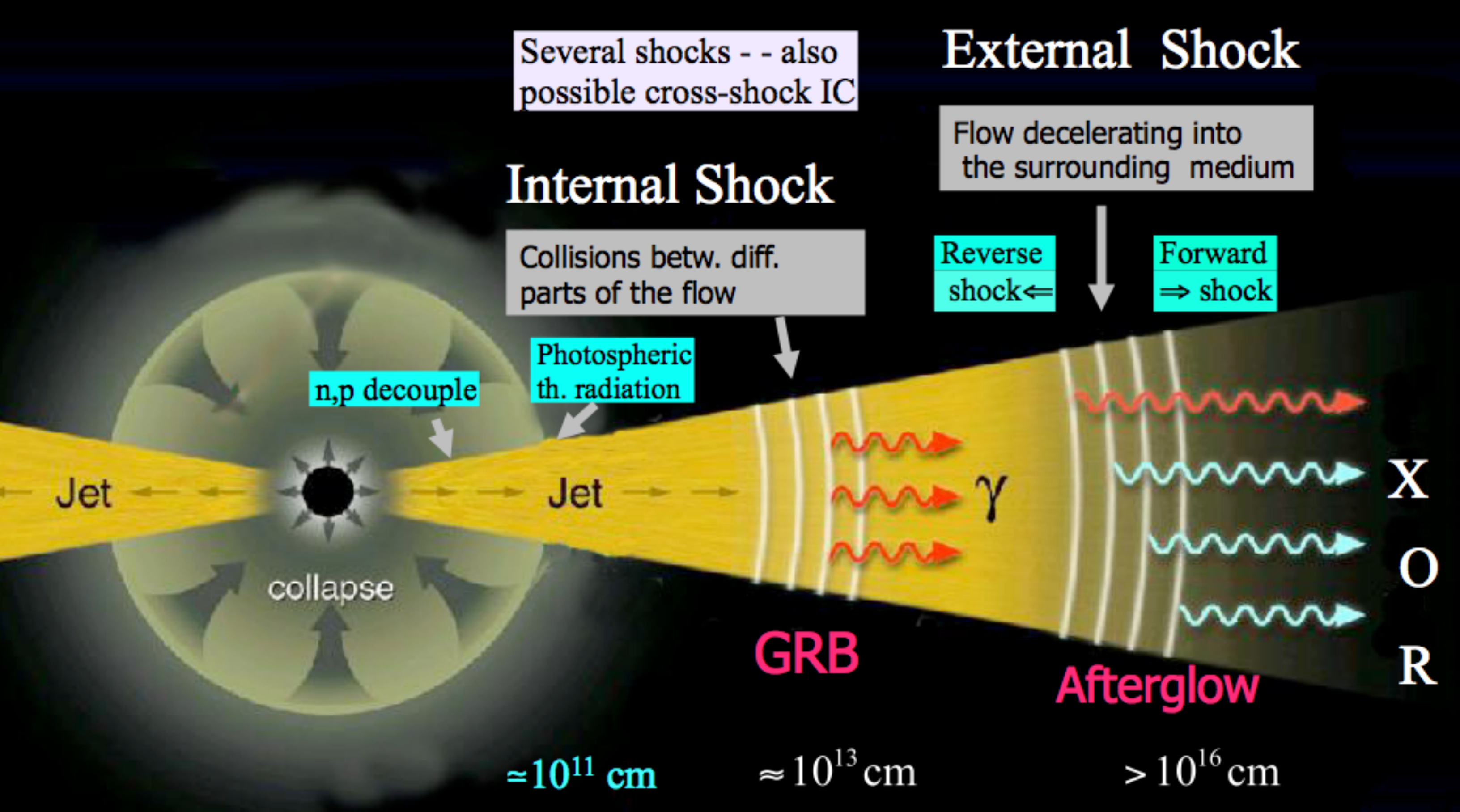}
  \caption{GRB jet shocks and emission regions, after \cite{Meszaros01sci}.}
\label{fig:jetshock}
\end{figure}

Such systems, consisting of as much as three different emission zones, the external shock 
furthermore consisting of both a forward shock and a reverse shock propagating onto the ejecta,
can give rise to a  variety of non-thermal spectra. Upscattering by electrons in the
various regions of photons arising in the same region or inner regions can give rise to
multiple radiation components, e.g. \cite{Meszaros06grbrev}.

\section{Observational progress through Swift and Fermi}
\label{sec:swifer}

The observations from the Swift and Fermi satellites, e.g. \cite{Gehrels+12sci,Omodei+13fermigrb}, 
have led to a number of interesting results. Swift  was able for the first time to follow GRB 
afterglow light curves  in X-ray and UV/O starting less than a minute after the $\gamma$-ray
trigger. This led, for instance, to the discovery in many afterglows of sudden drop of the X-ray 
light curves at the end of the prompt $\gamma$-ray emission, followed by a shallow decay lasting 
as much as 500 s in some cases, after which the canonical power law fireball decay took over 
(Fig. \ref{fig:lgrbsgrbz}, left). This behavior appears both in long and short GRBs, as expected 
from afterglows.
%
\begin{figure}[h]
\vspace*{-0.0in}
\begin{minipage}{0.5\textwidth}
\centerline{\includegraphics[width=3.1in]{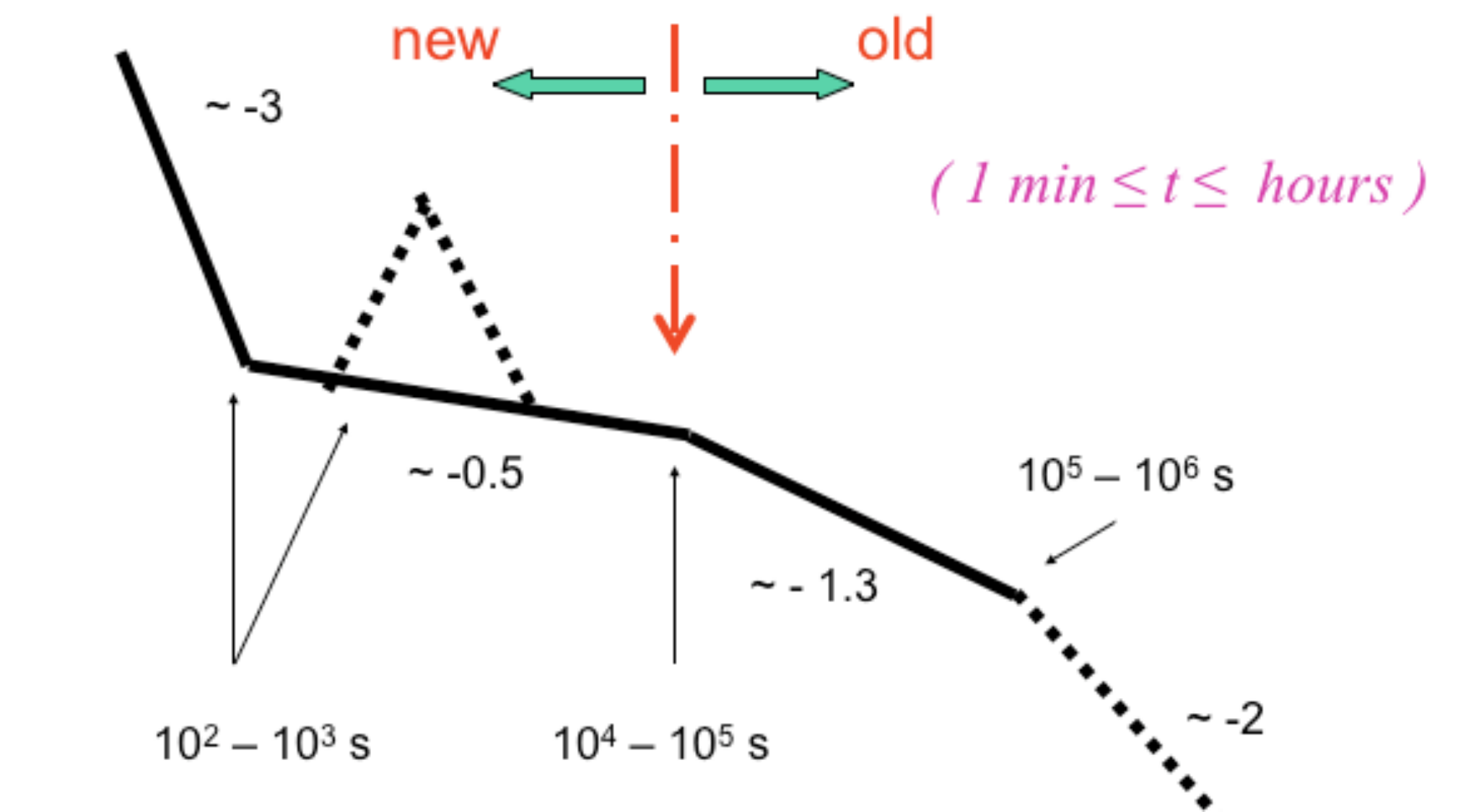}}
\end{minipage}
\begin{minipage}[t]{0.5\textwidth}
\vspace*{-1.2in}
\centerline{\includegraphics[width=3.1in]{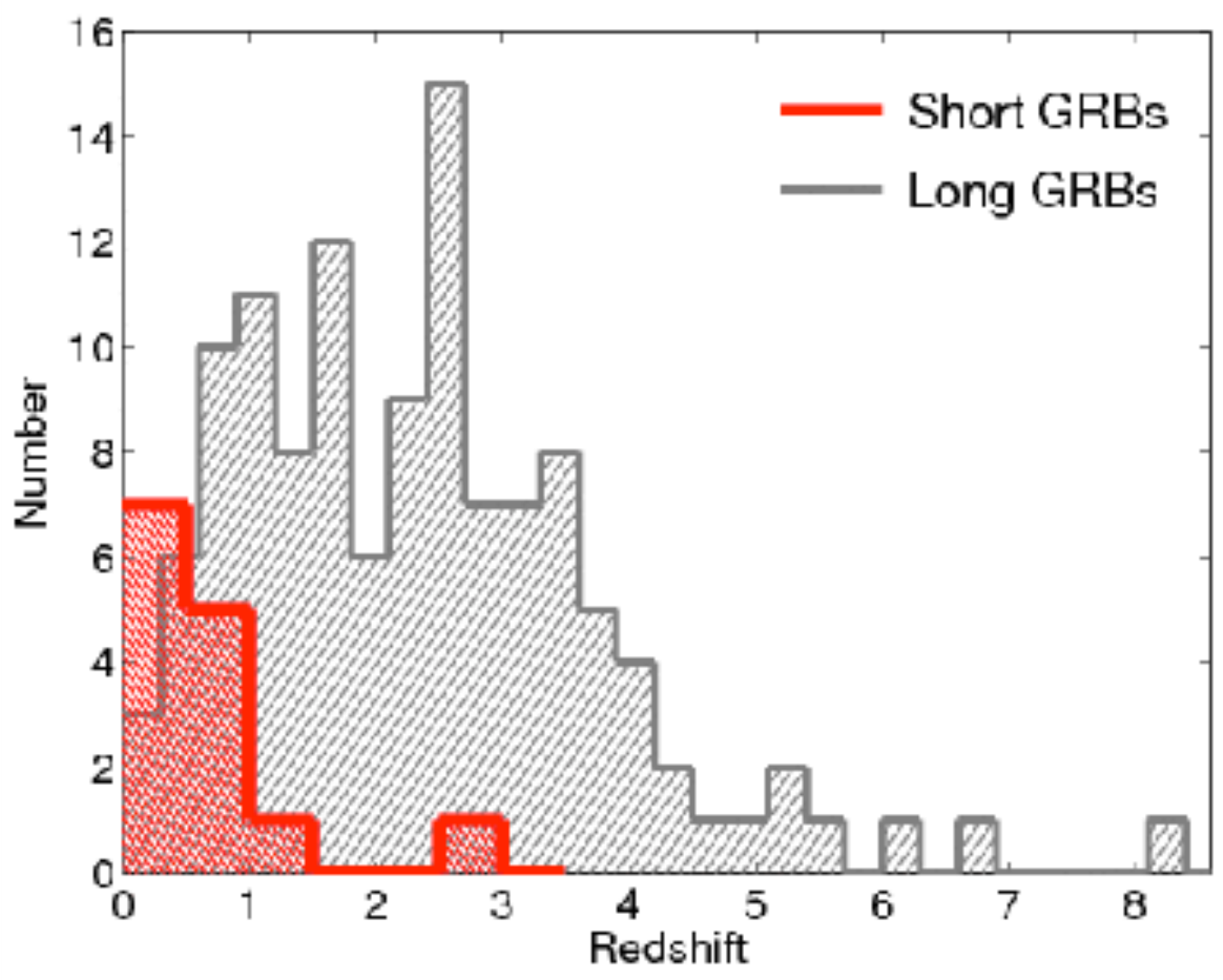}}
\end{minipage}
\caption{Left: schematic X-ray afterglow from Swift-XRT observations \citep{Zhang+06ag}.
Right: redshift distribution of long and short GRB (after E. Berger, 2014).}
\label{fig:lgrbsgrbz}
\end{figure}
The sudden drop may be due to the high latitude emission of the jet, while the shallow decay
may be due to continued outflow, with occasional X-ray flares (as seen in the figure) due to
late internal shocks, e.g. \cite{Zhang+06ag}.

Swift also made possible an explosive growth in the number of ground-based afterglow localizations 
with redshift determinations (see Fig. \ref{fig:lgrbsgrbz}, right), a smaller number of LGRBs having 
already been detected earlier by Beppo-SAX and HETE-2. Swift, in addition, localized for the first
time the host galaxies of short GRBs, which are detectable at redshifts typically lower then
LGRBs, and confirmed that the hosts of SGRBs include many old population (elliptical) galaxies,
compatible with an origin in compact double degenerate binaries, e.g. \cite{Nakar07sgrb,Gehrels+09araa}.
A significant fraction of short bursts, after their (previously known) initial short hard 
gamma-ray emission, also showed a previously unseen longer ($\sim$ 100 s) softer tail in their
light curve. Another surprise was the first detection of a shock breakout in a supernova
accompanied by a low luminosity soft GRB or X-ray flash, e.g. \cite{Campana+06-060218,Waxman+07-060218}.
The range of redshifts detected for LGRBs was dramatically extended, the highest spectroscopic
redshift so far being $z=8.2$, with a record  photometric redshift for GRB090429B at $z\simeq 9.4$,
e.g. \cite{Meszaros+12raa,Salvaterra15hizgrb}, see Fig. \ref{fig:lgrbsgrbz}, right.

Fermi, with its GeV range LAT  and MeV range GBM instruments, showed that many, if not most, burst 
detected with the LAT show that the GeV radiation starts to arrive seconds after the MeV trigger,
an example being shown in Fig. \ref{fig:080916-lc}. On the other hand, the GeV emission sometimes
lasts hundreds of seconds longer than the MeV, suggesting association with the afterglow. 
Various purely leptonic astrophysical mechanisms may give rise to this behavior, e.g., 
\cite{Kumar+14grbrev}. In addition, this behavior can be used also to provide interesting constraints 
on quantum gravity theories, e.g.  \cite{Abdo+09-090510-liv}.
\begin{figure}[htb]
\begin{minipage}{0.8\textwidth}
\includegraphics[width=1.0\textwidth,height=3.0in,angle=0.0]{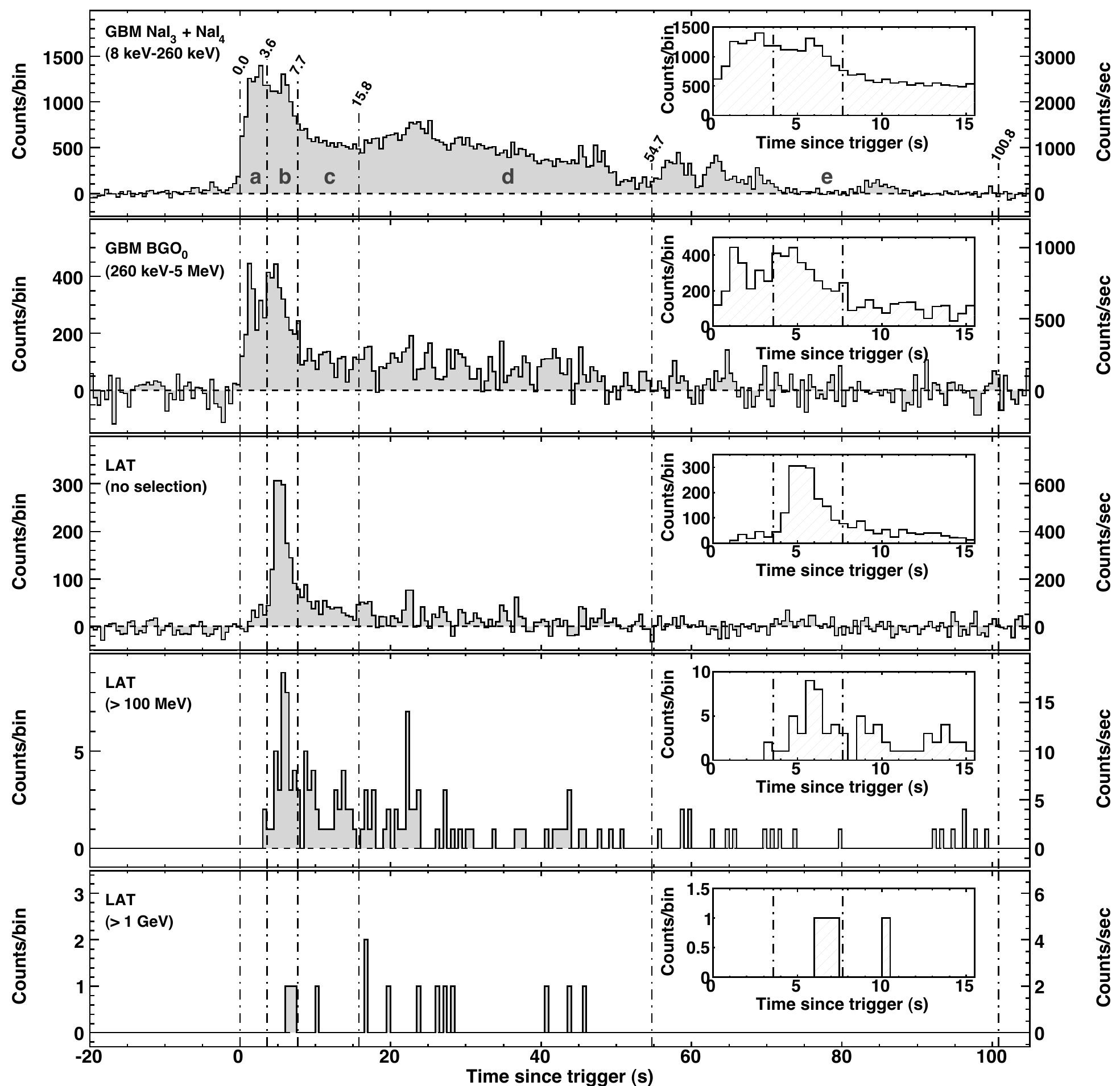}
\end{minipage}
\hspace{2mm}
\begin{minipage}[t]{0.17\textwidth}
\vspace*{-1.5in}
\caption{Light curves of GRB080916C showing the \gbm (top two curves) and
the \lat (bottom three curves) energy ranges  \cite{Abdo+09-080916}.}.
\end{minipage}
\label{fig:080916-lc}
\end{figure}

Another interesting although not unexpected  feature is the presence, in many LAT bursts,
of a second spectral component at high (GeV) energies, whose slope is harder than the
MeV classical broken power law (Band) spectral shape. More interesting was the fact that
in some LAT bursts a second separate component is not statistically significant, even though
after a few seconds the high energy slope of the Band function (which initially is soft and
does not extend to GeV) hardens and starts to extend into the GeV range. However, in a larger
number of LAT bursts a second component with a harder slope does appear, with some delay 
relative to the softer MeV Band function, see Fig. 5. 
Some theoretical interpretations of these features are discussed in the next section.
\begin{figure}[htb]
\begin{minipage}{0.75\textwidth}
\includegraphics[width=1.0\textwidth,height=2.0in,angle=0.0]{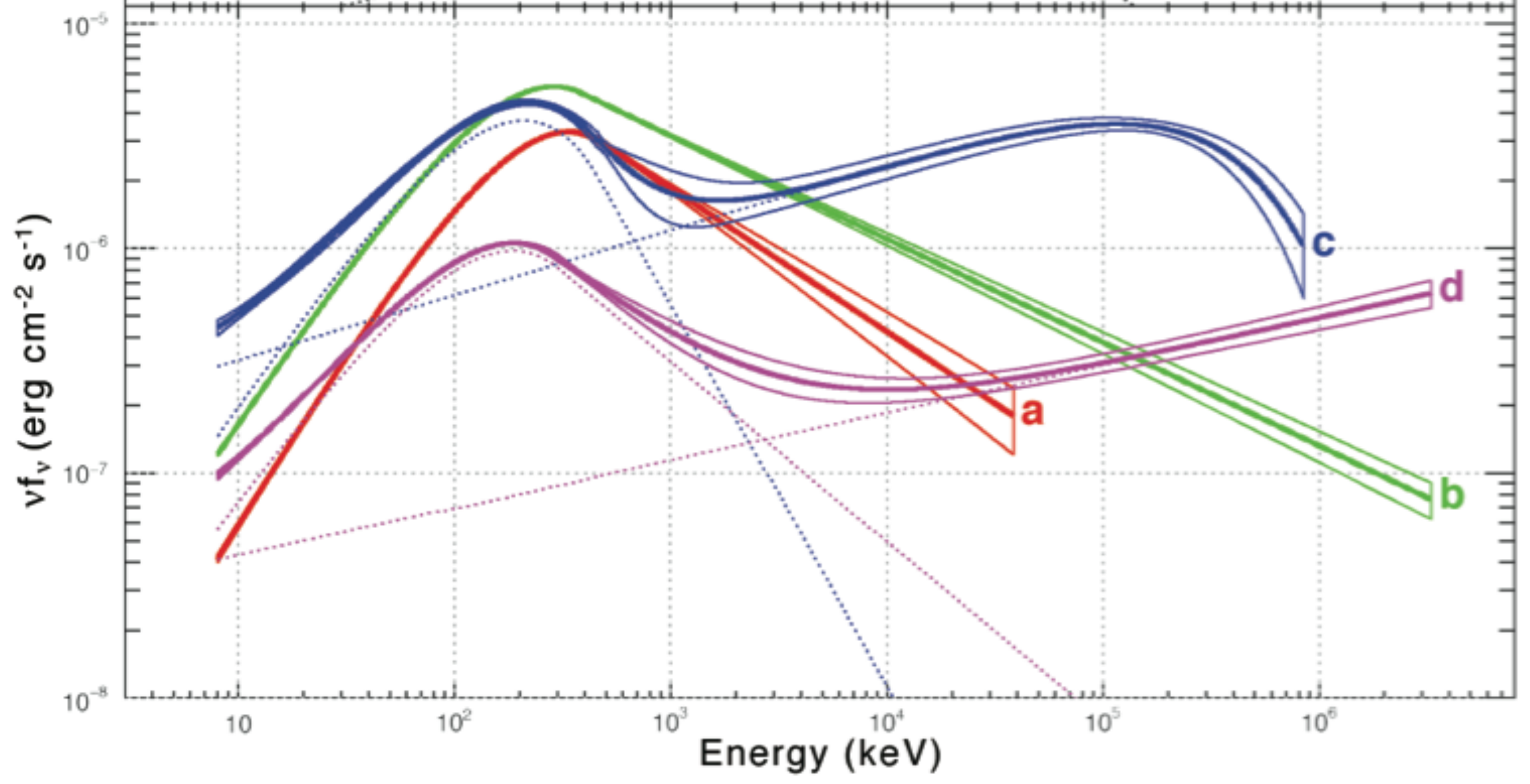}
\end{minipage}
\hspace{5mm}
\begin{minipage}[t]{0.2\textwidth}
\vspace*{-1.0in}
\caption{Spectra of GRB090926A from \fermi at four different time intervals,
a= [0.0-3.3s], b= [3.3-9.7s], c= [9.7-10.5s], d= [10.5-21.6s]
\cite{Ackermann+11-090926}.}
\end{minipage}
\label{fig:090926A-spec}
\end{figure}

More recently, a new class of ultra-long GRBs is being recognized. Some of them appear not 
be connected to ``usual" GRBs but appear rather to be tidal disruption events (TDEs)
of a normal star by a massive black hole at the center of a galaxy. The most prominent
example of such a TDE detected in $\gamma$-rays by Swift is the long-lasting source Sw J1644+57
\citep{Burrows+11-1644tid}, where the tidal disruption seems to produce a relativistic
jet leading to $\gamma$-ray, X-ray, optical and radio emission somewhat similar to a
scaled-up and stretched GRB. Other TDEs have been detected in X-rays or optical,
e.g. \cite{Komossa+15tid}, leading to an increased interest in TDE model calculations,
e.g. \cite{Liu+12sw1644tid,Stone+13tid,Coughlin+14tid,Kelley+14tid,Liu+15tid,Mimica+15tid,
Piran+15tid,Piran+15tidcirc}.

However some ultra-long GRBs (loosely defined as burst-like objects lasting upwards of 1000 s 
in gamma-rays) may be due to a longer version of stellar events similar to classical GRBs,
e.g. \cite{Gruber+11ulgrb091024,Stratta+13ulgrb111209,Levan+14ulgrb111209}.
One way in which such ultra-long gamma-ray bursts (ULGRBs) could arise is if the progenitor
is larger than the usual Wolf-Rayet progenitor (WRs lost through winds their outer envelope 
and have radii $\sim 10^{10}-10^{11}\cm$ so jet breakout plus accretion times accord with 
the classical LGRB durations, e.g. \cite{Woosley+99engine}). 
Instead, ULGRBs might arise in blue supergiants, which have an 
extended envelope, or perhaps even in ultra-massive Pop. III (first generation) stars,
e.g. \cite{Gendre+13ulgrb111209,Virgili+13ulgrb091024,Zhang+14ulgrb,Evans+14ulgrb130925,
Boer+15ulgrb,Suwa+11grbpop3,Nakagura+12pop3,Yoon+15ulgrbpop3,Gao+14ulgrb}. Early jet breakout 
calculations from blue supergiant progenitors were discussed in \cite{Meszaros+01jetbubble}, 
while ultra-long duration X-ray and O/IR signatures of Pop. III stars were estimated in,
e.g. \cite{Komissarov+10pop3,Meszaros+10pop3,Toma+11pop3}, and the possibility of testing for 
an extended envelope or a Pop III progenitor through a high energy neutrino precursor 
was considered in \cite{Razzaque+03nutomo,Schneider+02pop3grbnu, Gao+11pop3nu}.

\section{Some recent theoretical developments}
\label{sec:theo}

The afterglow model based on an external forward and reverse shock is, on the whole,
able to explain the basic features of the emission in the latter phases.  The internal shock 
model, on the other hand, which is widely used by observers as the work-horse tool for
interpreting the data, in its simple form has two problems: one is a low radiative efficiency,
due to  the low mechanical shock efficiency, and the other its asymptotic theoretical low energy 
spectrum, which for some bursts is softer than observed. Two main avenues have been explored
in the past decade to resolve this issue. 

\begin{figure}[htb]
\begin{minipage}{0.5\textwidth}
\includegraphics[width=1.0\textwidth,height=2.5in,angle=0.0]{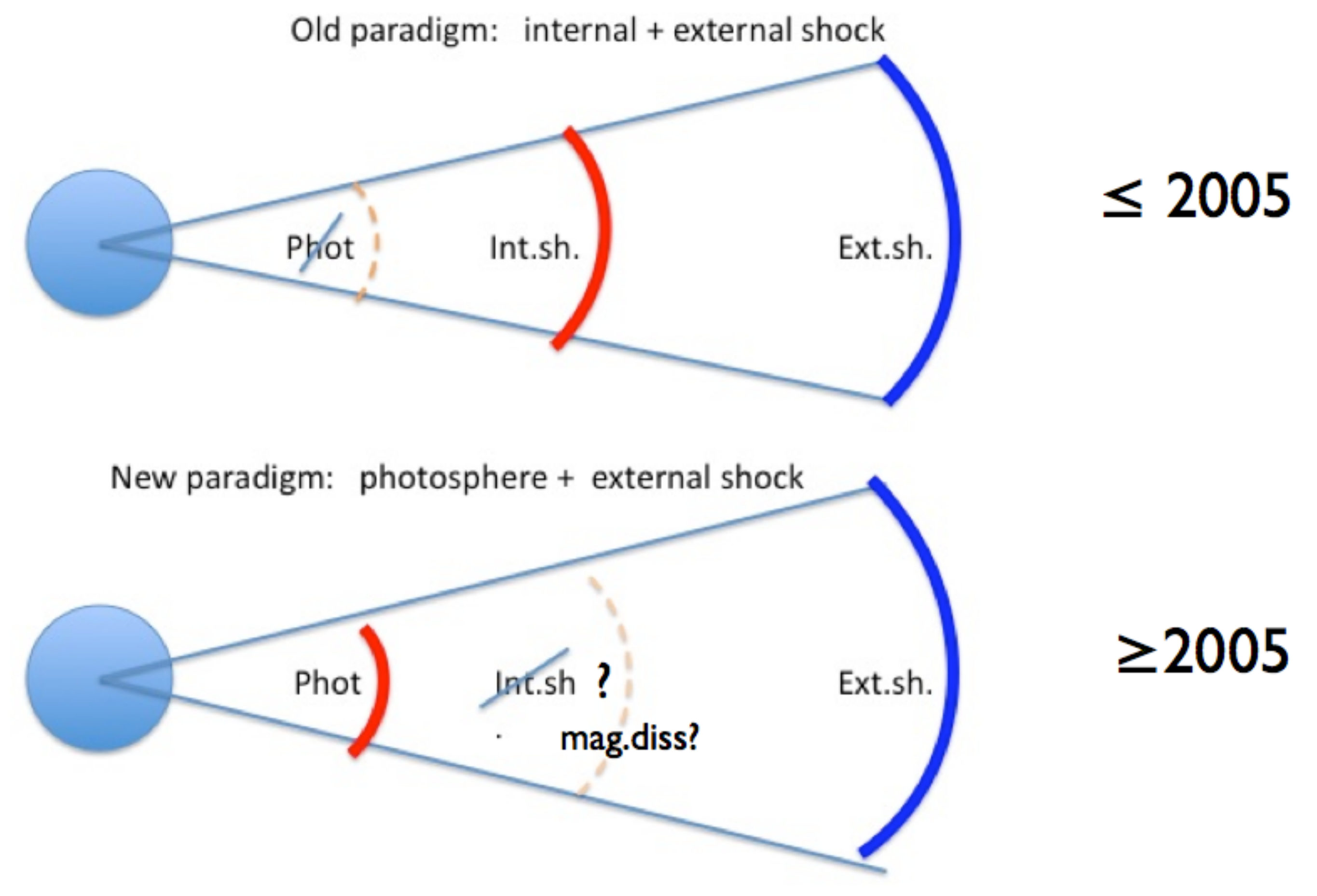}
\end{minipage}
\hspace{5mm}
\begin{minipage}[t]{0.3\textwidth}
\vspace*{-1.0in}
\caption{One version of the evolving paradigm for GRB emission, where the prompt MeV
emission is attributed to a dissipative photosphere where non-thermal processes
modify the thermal spectrum.}
\end{minipage}
\label{fig:phot-paradigm}
\end{figure}
One way to address this is to consider, instead of the usual non-dissipative (and hence 
inefficient) photosphere plus internal shock, a new paradigm consisting of a photosphere 
which has a hard low energy slope, with nonthermal high energy contributions from Comptonization
and/or additional internal shocks \citep{Meszaros+00phot}. The photosphere can be radiatively
efficient if it has internal dissipation, e.g. from sub-photospheric shocks \citep{Rees+05phot},
and there would still be also an external shock as well, see Fig. \ref{fig:phot-paradigm}.
Numerical simulations  \citep{Peer+06phot} including sub-photospheric shocks followed by synchrotron 
and inverse Compton scattering, i.e. a leptonic mechanism, confirm both a high efficiency and 
reasonable low and high energy slopes as well as break energies.

Another source of sub-photospheric dissipation is inevitable, of the outflow baryonic load 
contains both protons and neutrons, which should be the case. This is because the $pn$ fluid,
while initially expanding together well coupled by elastic collisions, eventually decouple,
the neutrons lagging behind, which leads to inelastic collisions resulting in pions, whose
decay produces both neutrinos and relativistic positrons \citep{Derishev+99,Bahcall+00pn}.
This also results in efficient dissipation, the non-thermal positrons heating the plasma
and leading to escaping photon spectra which again are reasonable \citep{Beloborodov10pn},
see Fig. 7.  
\begin{figure}[htb]
\begin{minipage}{0.5\textwidth}
\includegraphics[width=1.0\textwidth,height=2.5in,angle=0.0]{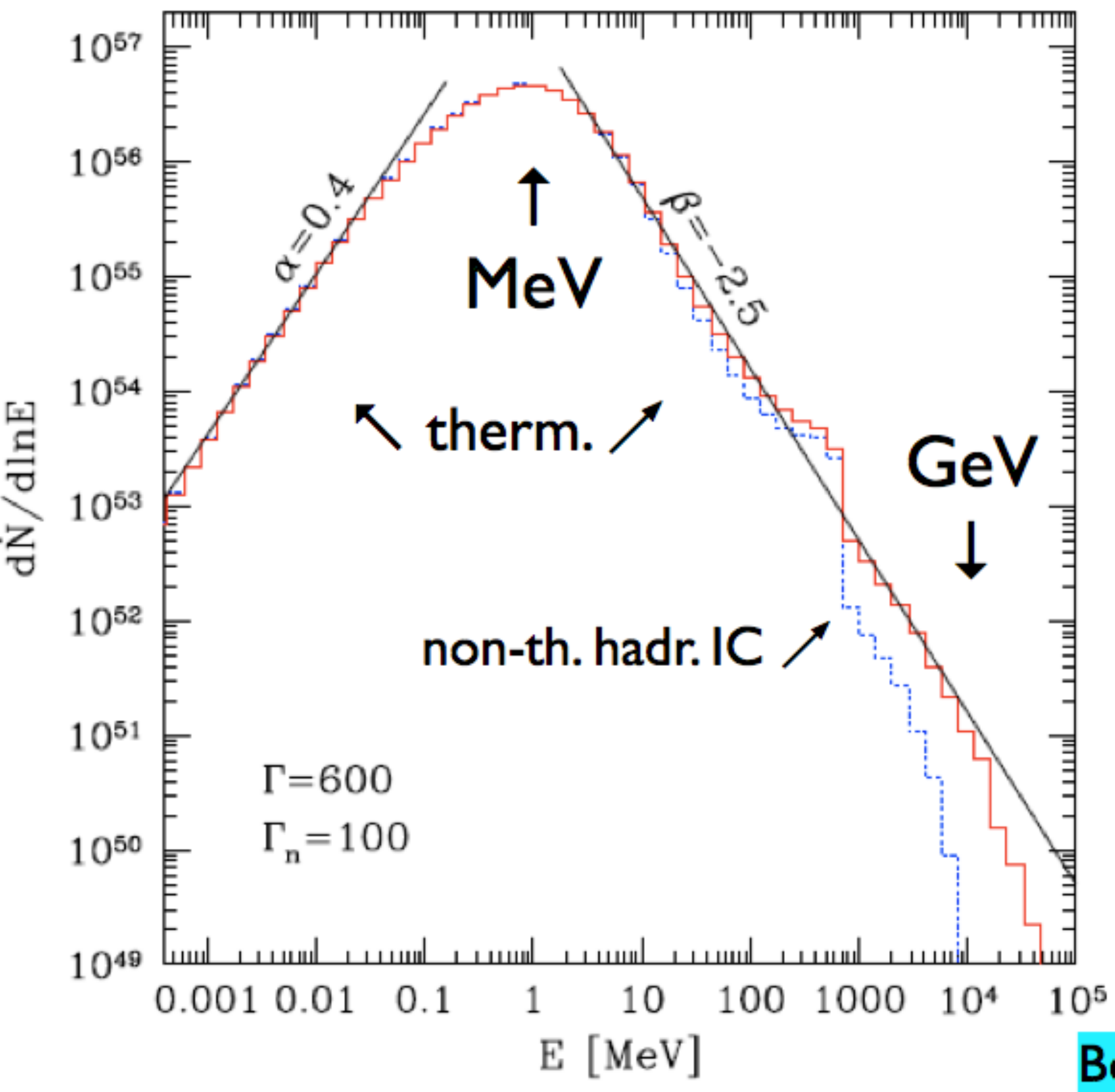}
\end{minipage}
\hspace{5mm}
\begin{minipage}[t]{0.3\textwidth}
\vspace*{-1.0in} 
\caption{Dissipative photosphere heated by $np$ collisions leading to a non-thermal
``Band" spectrum \citep{Beloborodov10pn}. The high energy slope extends into the GeV
band thanks to energy injection from the non-thermal decay positrons. In this case
it reproduces a single Band function, but adding magnetic fields can add a second
high energy component with a harder slope \citep{Vurm+13phot}.}
\end{minipage}
\label{fig:phot-belo10}
\end{figure}

A more robust way to obtain a second high energy component in such dissipative photospheric
models is through the up-Comptonization of the inner photospheric spectrum by the external
forward and reverse shock \citep{Veres+12mag}, see Fig. \ref{fig:barphot}.{
%
\begin{figure}[h]
\vspace*{-0.0in}
\begin{minipage}{0.5\textwidth}
\centerline{\includegraphics[width=3.1in]{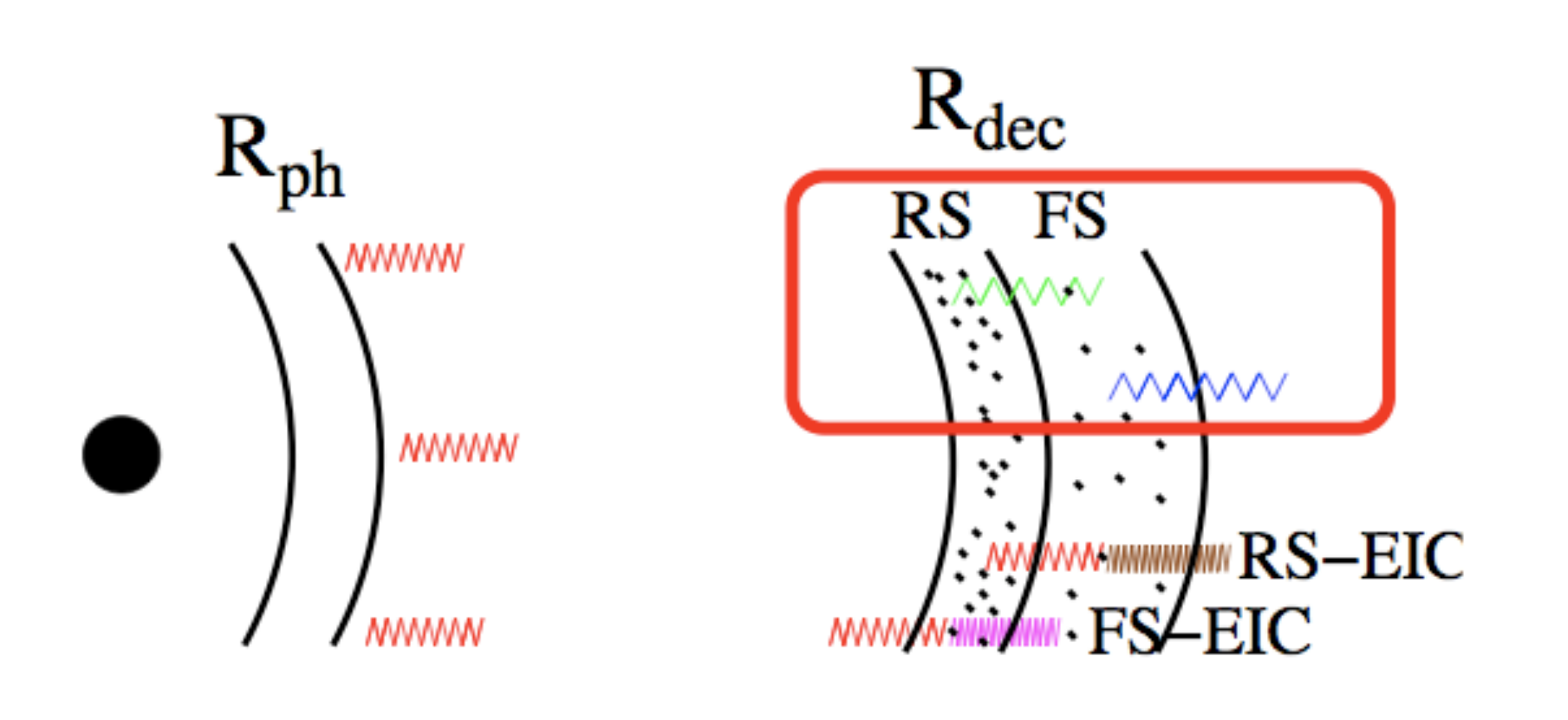}}
\end{minipage}
\begin{minipage}[t]{0.5\textwidth}
\vspace*{-1.00in}
\centerline{\includegraphics[width=3.1in]{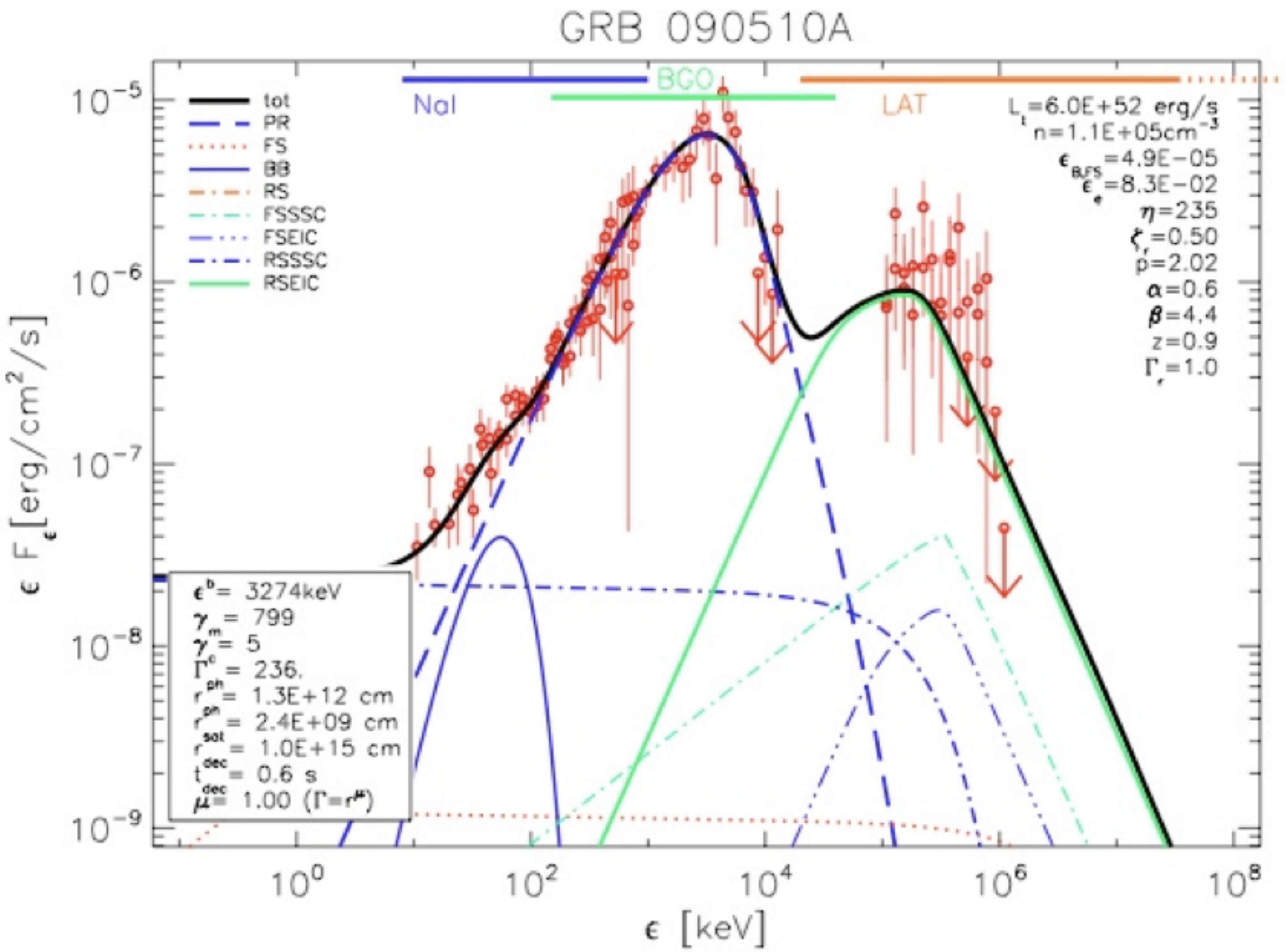}}
\end{minipage}
\caption{Left: schematic dissipative photosphere and upscattering by external shock. 
Right: baryonically dominated dissipative photosphere plus external shock upscattering
model leading to a two-component Band plus GeV spectrum \citep{Veres+12mag}.}`
\label{fig:barphot}
\end{figure}

The other main way to address the efficiency and spectral issues of internal shocks is by including 
the acceleration of hadrons in the same internal shocks,  which is ignored in the classical internal 
shock picture, but which should be essentially unavoidable, if the outflow is baryonic. 
Hadron acceleration in internals shocks was first considered by \cite{Waxman+97grbnu} to predict
TeV neutrinos from GRBs, and the associated high energy $\gamma$-rays were considered by various
authors, e.g. \cite{Boettcher+98hadrgrb,Asano+09grb}. The high energy components are produced
by the synchrotron and inverse Compton of the cascade secondaries, see Fig. \ref{fig:ppxsec}, left.
The delay associated with the acceleration and cascade development can also possibly explain the 
delay of the GeV emission relative to the MeV radiation \citep{Asano+09grb,Razzaque10-090510,Asano+13phot}.
%
\begin{figure}[h]
\vspace*{-0.0in}
\begin{minipage}{0.5\textwidth}
\centerline{\includegraphics[width=3.1in]{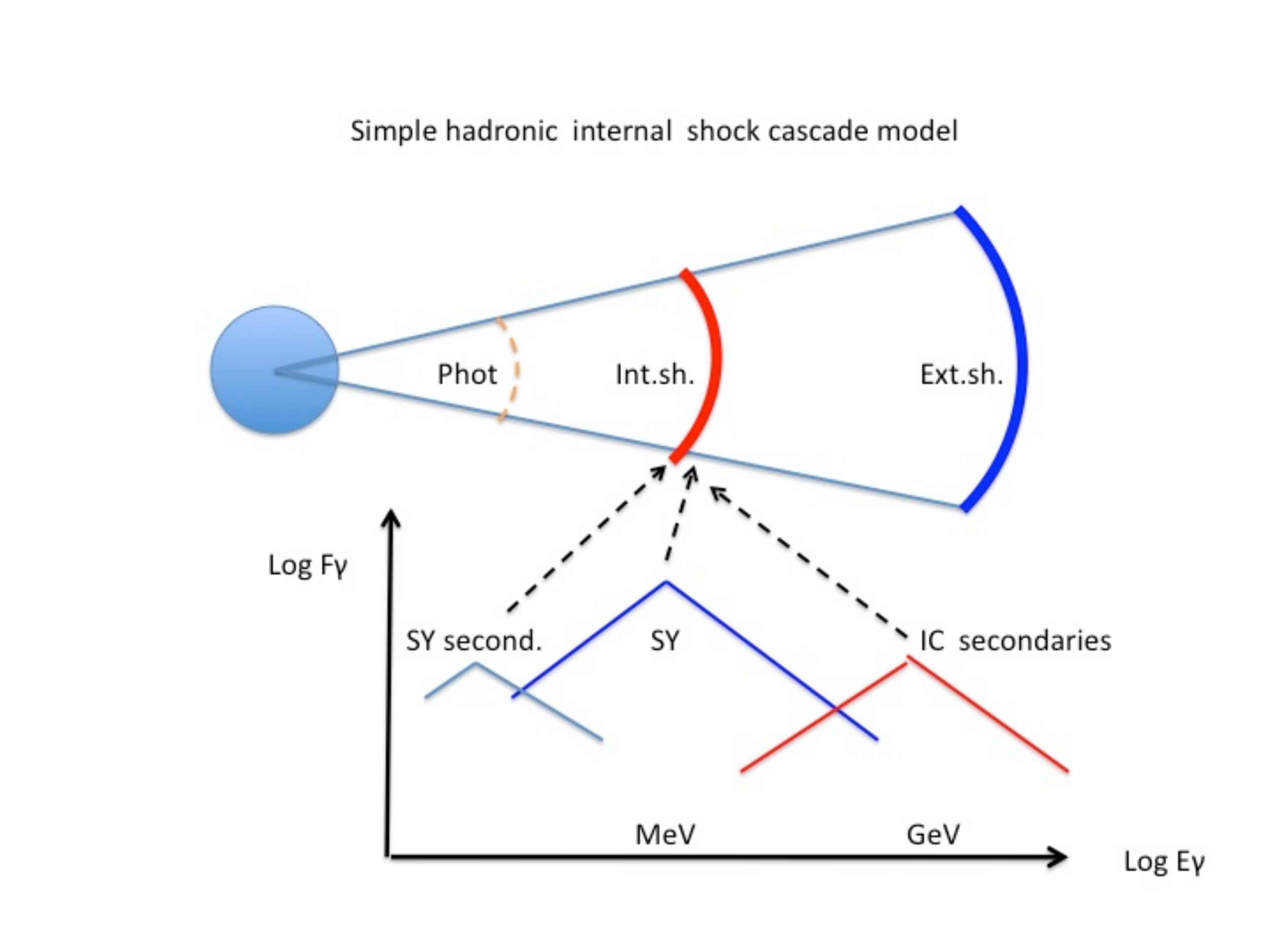}}
\end{minipage}
\begin{minipage}[t]{0.5\textwidth}
\vspace*{-1.2in}
\centerline{\includegraphics[width=3.1in]{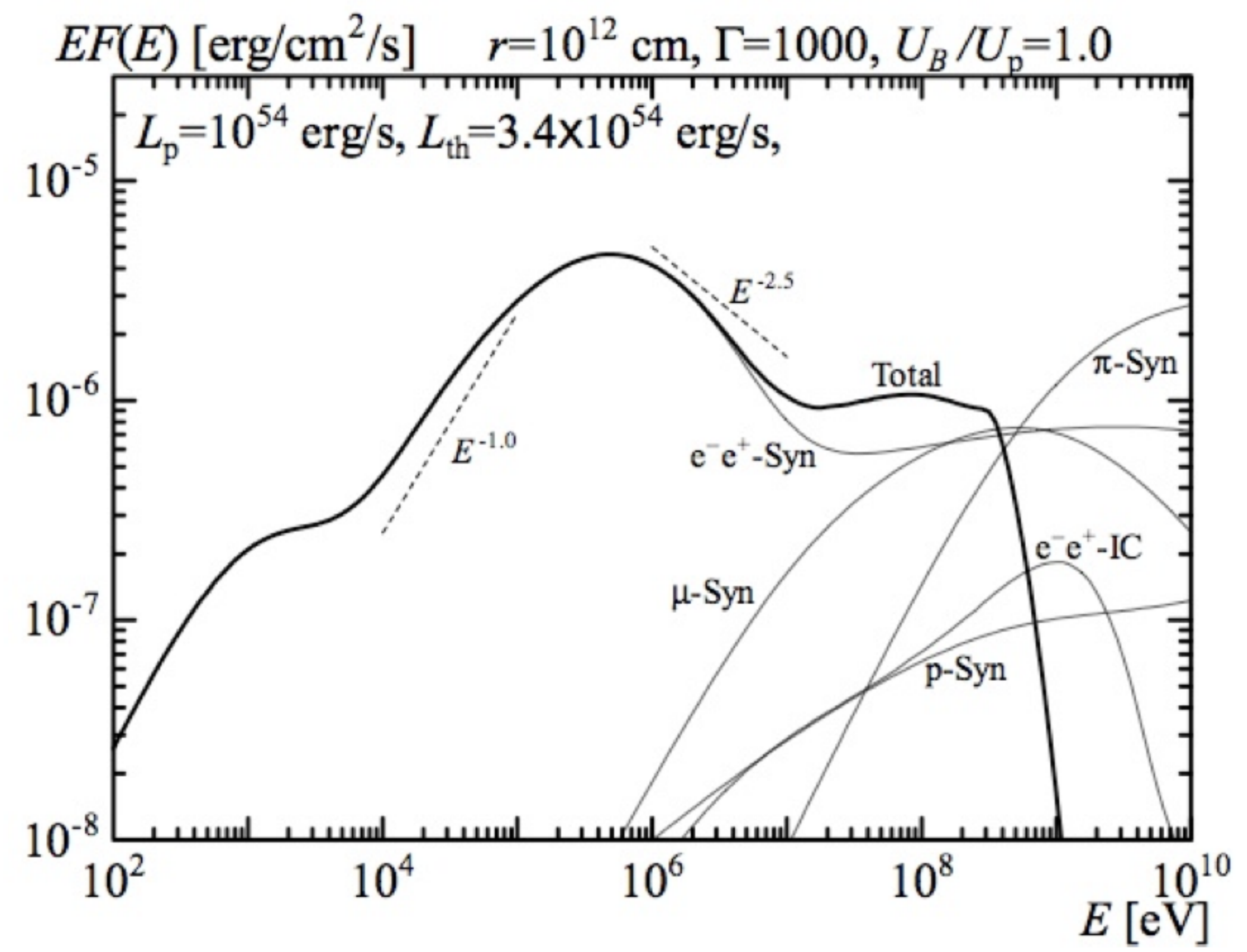}}
\end{minipage}
\caption{Left: schematic spectrum internal shock with electron plus hadron acceleration and 
secondary cascades.  Right: numerical calculation of hadronic internal shock leptonic plus 
cascade secondary spectrum \citep{Murase+12reac}}.
\label{fig:ppxsec}
\end{figure}

The hadronic secondaries can reproduce both the high energy second component, and also an additional 
low energy component which appears associated with the high energy component. This is illustrated
as the leftmost component in Fig.  \ref{fig:ppxsec} left, a hint of which is seen in the low energy 
portion of the data of Fig. 5, and more clearly in GRB090510, a calculation of which is given
in \cite{Asano+09-090510}. This component could also give rise to the bright early optical flash 
seen in some bursts.  An interesting possibility is that re-acceleration of the secondaries via
Fermi 2nd order mechanism in the turbulence behind the internal shocks can lead to a self-consistent
Band spectrum at MeV energies, see Fig. \ref{fig:ppxsec} (right), which thus may not need to
be due to the photosphere. The radiative efficiency in this case is also high, as well as the
spectrum being reasonable, without an unduly high proton luminosity.

\section{Multi-messenger approaches}
\label{sec:multi}

The multi-messenger signatures of GRB are an important direction being actively pursued through
multiple channels, both photonic and non-photonic. The latter, e.g. in the form of neutrinos,
cosmic rays and gravitational waves, are being pursued by multiple observatories, such as
IceCube, Auger, Telescope Array, LIGO and VIRGO, in addition to VHE $\gamma$-ray facilities such as
HAWC, VERITAS, HESS and others, complementing Fermi, Swift and other existing instruments.
An ambitious effort to coordinate the sub-threshold signals from various such facilities is
the AMON network (http://amon.gravity.psu.edu, see \cite{Smith+12amon}).
An interesting question is, if GRBs accelerate hadrons, can they explain at least partly the
Auger CR and IceCube neutrino observations, or do they satisfy the constraints imposed by these
observations? And will short GRBs,  if they are indeed merging neutrons star binaries, be detected
in the near future by gravitational wave observatories?

With IceCube and Auger in full operation, the first question is being already addressed. 
Auger has measured the ultra-high energy cosmic ray (UHECR) spectrum to great precision 
\citep{Auger+13icrc33}, but since CRs are deflected by intergalactic and galactic magnetic fields, 
one can only test whether GRBs can produce  the appropriate flux within the GZK radius, as 
suggested by \cite{Waxman95cr}. Since the GRB UHECRs would produce via $p\gamma$ TeV-PeV
neutrinos \citep{Waxman+97grbnu}, these should not exceed the IceCube limits. 
The first limits were set by the IceCube initial 40+59 string array \citep{Abbasi+12-IC3grbnu-nat},
which compared neutrino observations at the locations and time windows corresponding to 300 GRBs
detected electromagnetically by satellites, 183 GRBs with the 50 string array and 177 GRBs with 
the 40 string array.  The comparison used the simplified internal shock photo-hadronic model 
of \cite{Waxman+97grbnu} to predict an expected neutrino flux, resulting in an 
over-prediction of this model compared to the data by a factor $\sim 5$. A more generic model 
independent test, where the detected Auger cosmic rays are attributed to the GRBs was also negative.
However, more accurate descriptions of the photo-hadronic interactions as well as inclusion of 
astrophysical model uncertainties \citep{He+12grbnu,Li12ic3nu,Hummer+12nu-ic3} showed that 
discrepancies between this model and IceCube might only be expected after several years of 
observations with the full array.  These tests also used stationary radius internal shocks, 
allowing at most for range of radii (through $t_v$). 

\begin{figure}[ht]
  \centering
  \includegraphics[width=14 cm]{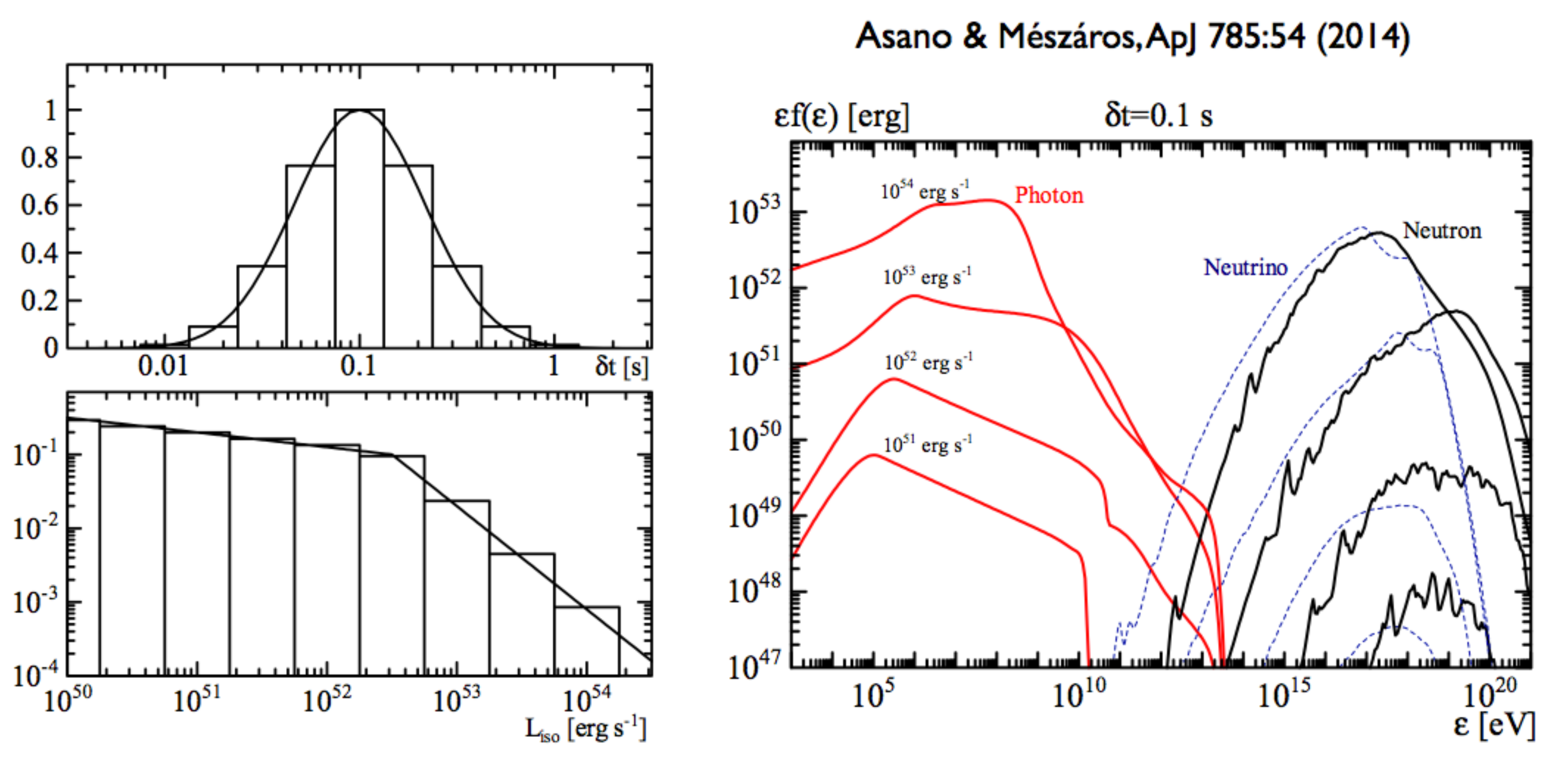}
  \caption{Left: observed distribution of variability times $t_v$ (left top) and GRB luminosity 
function (left, bottom, from \cite{Wanderman+10grbsfr}. 
Right: Time-dependent hadronic internal shock Monte Carlo cascade spectral calculation for a 
distribution of variability times and observed luminosity function, showing photon, neutron and 
neutrino spectrum  \citep{Asano+14grbcr}}.
\label{fig:asano14-f1f2}
\end{figure}
However, the internal shocks, after forming, continue
to move out, and doing so expand, so the time dependence of the varying comoving quantities
must be taken into account. Such a time-dependent calculation was performed by \cite{Asano+14grbcr},
who took into account also the observed range of $t_v$ as well as the shock expansion and the GRB 
luminosity function \citep{Wanderman+10grbsfr} to obtain the expected flux of $\gamma$-rays, 
neutrinos and neutrons escaping from the GRBs - see Fig. \ref{fig:asano14-f1f2} (right).

These sources were used then to calculate the expected diffuse fluxes, using the luminosity 
function of Fig. \ref{fig:asano14-f1f2} (left). These predicted diffuse fluxes are compared in 
Fig. \ref{fig:asano14-f3f4} to the IceCube 40+59 limits \citep{Abbasi+12-IC3grbnu-nat} and the 
diffuse Auger UHECR flux \citep{Auger+13icrc33}. 
Two prescriptions were used for the UHECR escape, one (Fig. \ref{fig:asano14-f3f4} left) assuming
that they escape as neutrons due to $p\gamma$ interactions and then reconvert to protons once
outside the source; the other extreme being simple free escape as protons 
(Fig.  \ref{fig:asano14-f3f4} right). It can be seen that such GRB time-dependent internal
shocks can explain at least part of the highest energy UHECR in the GZK range, without violating
the 40+50 string IceCube neutrino limits.
%
\begin{figure}[h]
\vspace*{-0.0in}
\begin{minipage}{0.5\textwidth}
\centerline{\includegraphics[width=3.1in]{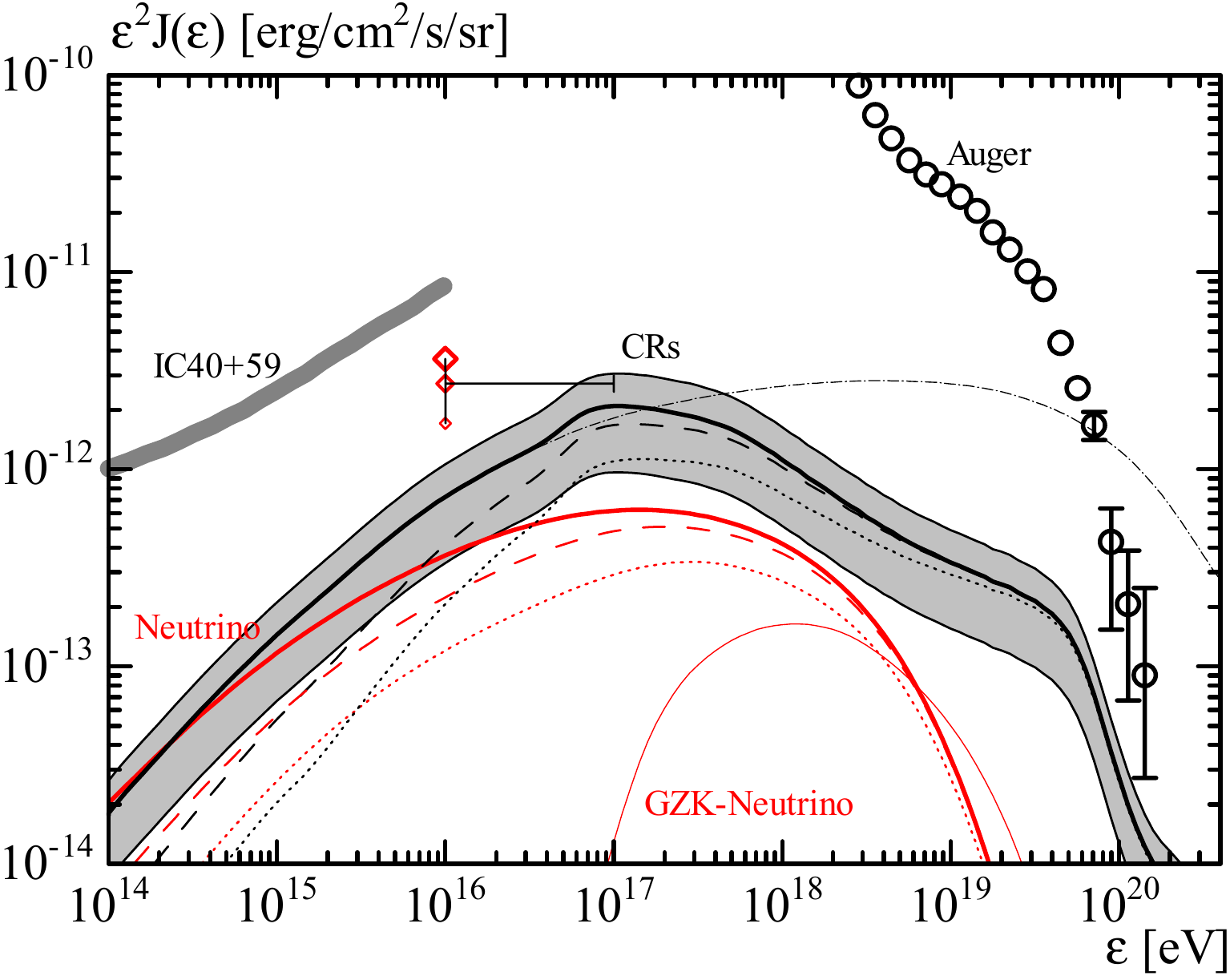}}
\end{minipage}
\begin{minipage}[t]{0.5\textwidth}
\vspace*{-1.2in}
\centerline{\includegraphics[width=3.1in]{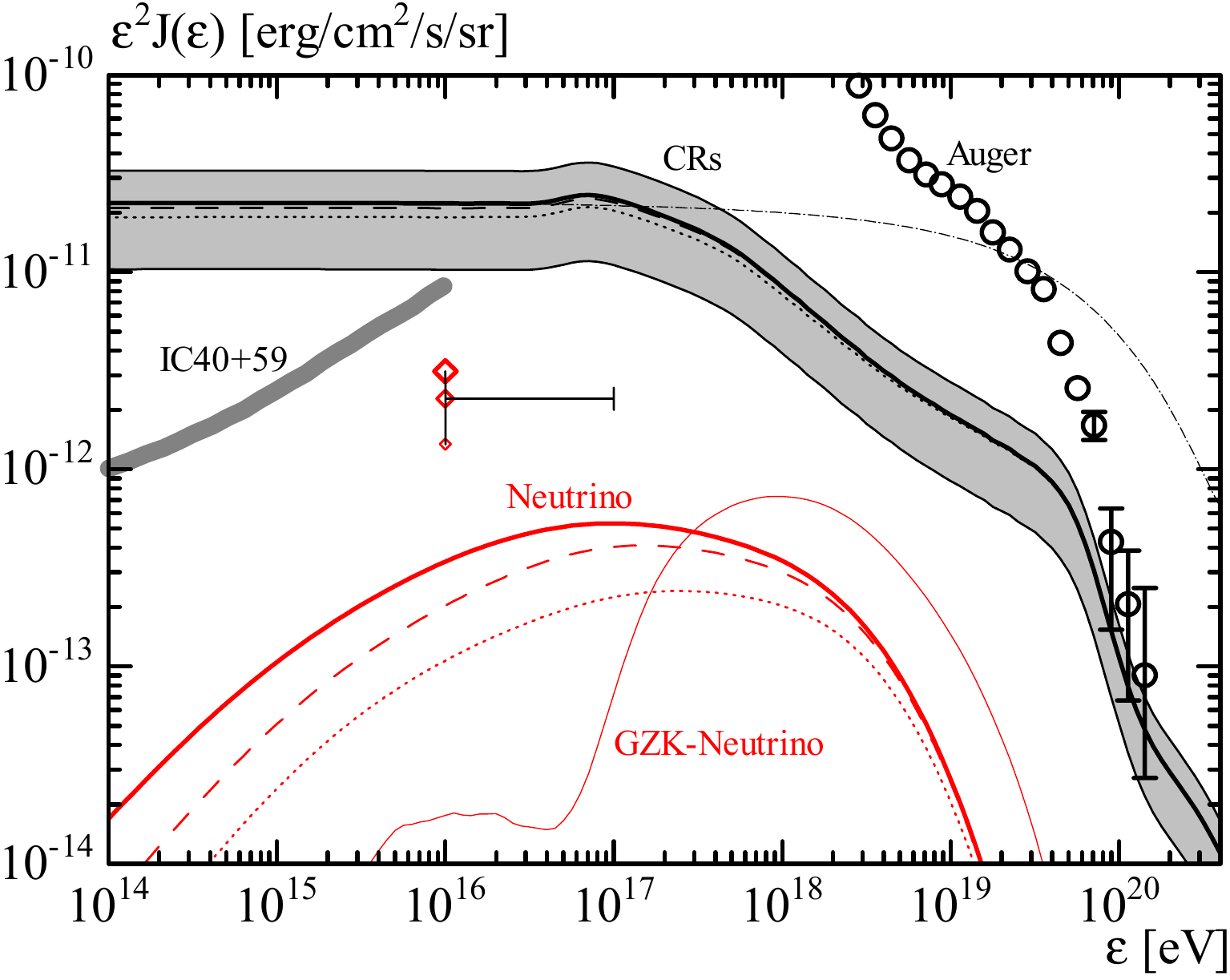}}
\end{minipage}
\caption{Left: the CR (black) and neutrino (red) diffuse intensities for the neutron conversion model,
compared to the Icecube 40+59 \citep{Abbasi+12-IC3grbnu-nat} and the Auger data \citep{Auger+13icrc33}.
Gray shaded area for CRs indicates the uncertainty in the local GRB rate. The thin dash-dotted line 
is the CR spectrum without the effects of photomeson production and Bethe--Heitler pair production.
The cosmogenic neutrino spectrum produced via the GZK process is also shown
as the thin red line.
Right: same as left, but for the sudden release model. From \cite{Asano+14grbcr}}.
\label{fig:asano14-f3f4}
\end{figure}

As mentioned earlier in \S \ref{sec:theo}, one of the new approaches to describing the prompt
GRB $\gamma$-ray emission ascribes it to a dissipative photosphere, with a diminished role for
the internal shocks (e.g. if the outflow is magnetically dominated, internal shocks might be
prevented). Thus, it is necessary to consider also the possible hadron acceleration and
neutrino production in the GRB photospheres \citep{Murase08grbphotnu,Wang+09grbphotnu,Gao+12photnu,
Asano+13phot,Zhang+13grbnu,Murase+13subphotnu,Bartos+13pn}. In addition, modified internal shocks
might also include magnetic dissipation, e.g. the ICMART model \citep{Zhang+11icmart}, and both these
and photospheric models predict qualitatively different neutrino fluxes than classical internal shocks.
A recent IceCube analysis, using 4 years of accumulated data including now the full array, 
tests such models \citep{IC3+15grbnu4yr}, concluding that less than 1\% of the electromagnetically
detected GRBs (modeled as either classical internal shocks, ICMART shocks or photospheres) can be 
contributing to the diffuse IceCube neutrino flux.

An exciting development is the discovery by IceCube of a diffuse neutrino flux which appears
to be of astrophysical origin \citep{IC3+13pevnu1,IC3+13pevnu2}, with limited (or no)
contribution from our own galaxy \citep{Fox+13pev,Ahlers+14pevnugal,Anchordoqui+14pevnugal,
Ahlers+15galpevnu}.  This flux has a spectrum extending from about a few PeV down to $\sim 10\TeV$, 
with a slope which (if unbroken) appears to be $\sim -2.3$ to $-2.5$.
As mentioned above, less than 1\% of these can be coming from ``classical", i.e. normal
high luminosity GRBs detected by Swift or Fermi \citep{IC3+15grbnu4yr}. In fact, looking at
Fig. \ref{fig:asano14-f3f4}, it is clear that such GRBs, while not violating the IceCube 
observational limits, in fact should not be producing anything near the observed neutrino
flux in the TeV-PeV range. The question then is, what are the sources of these PeV neutrinos?
If they are connected to GRBs at all, it might perhaps be to ``hidden" GRBs, which are
electromagnetically weak and do not trigger Swift or Fermi \citep{Murase+13choked,Liu+13pevnugrb}; 
these could be either so-called choked-jet GRBs \citep{Meszaros+01choked}, whose jets never 
emerged from the progenitor star, or they might be (electromagnetically) low-luminosity 
GRBS, or LLGRBs, e.g. \citep{Murase06llgrbnu,Gupta+07nullgrb,Murase+08llgrbcr}.

Other, non-GRB related possible sources of the diffuse TeV-PeV neutrinos could be galaxy
cluster accretion shocks and/or starburst galaxies, e.g. \cite{Loeb+06nustarburst,Thompson+06sbgnu, 
Murase+13pev,He+13pevnuhn,Anchordoqui+14pevnu,Tamborra+14igbnu,Chang+14pevnugam,Chang+15pevnugam}; 
or they could be due to AGN jets, e.g. \cite{Becker+14agnpevnu,Murase+14agnjetnu,Krauss+15agnpevnu}; 
or they could be due to galaxy-galaxy collisions \citep{Kashiyama+14pevmerg}.

\begin{figure}[htb]
\begin{minipage}{0.6\textwidth}
\includegraphics[width=1.0\textwidth,height=3.0in,angle=0.0]{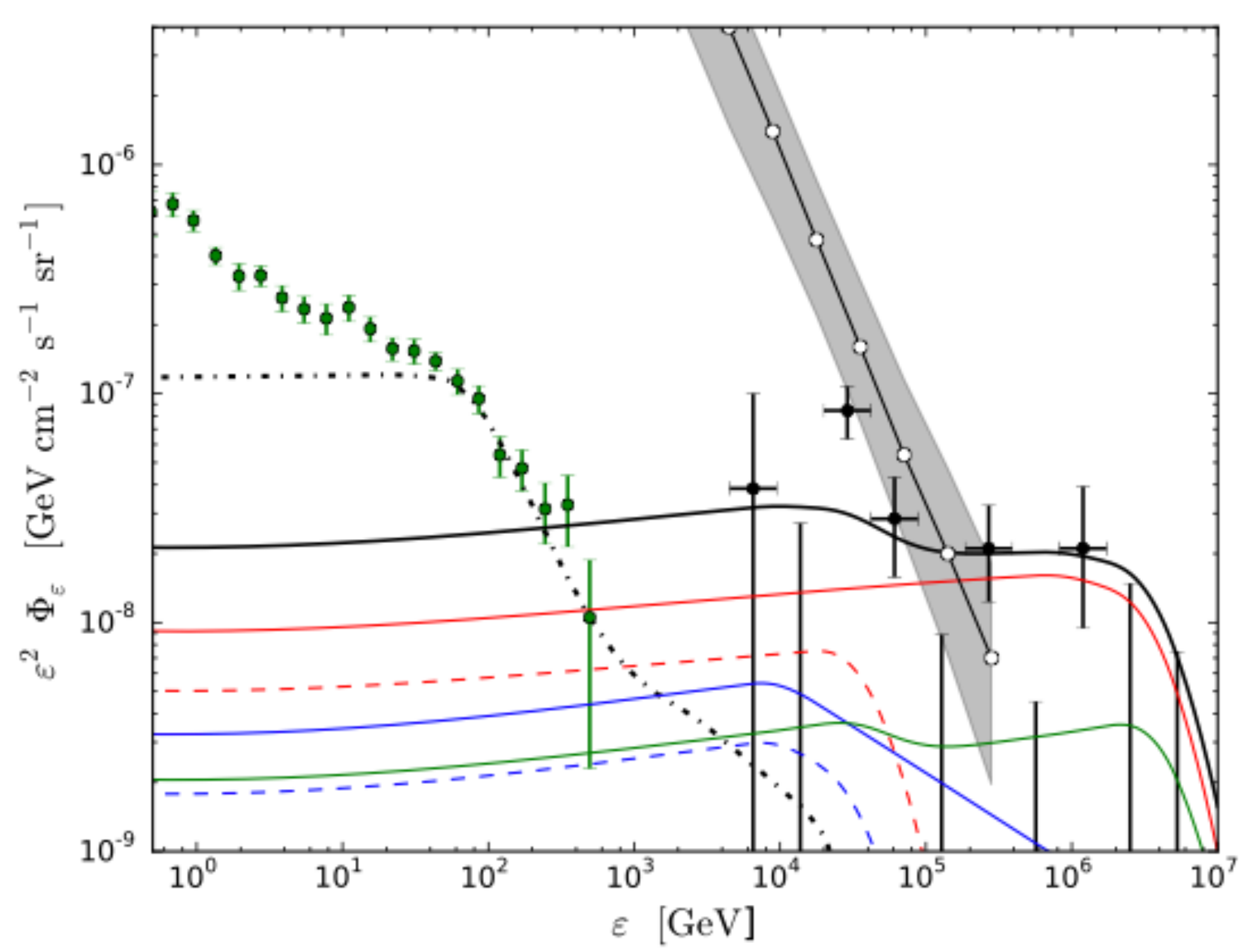}
\end{minipage}
\hspace{2mm}
\begin{minipage}[t]{0.37\textwidth}
\vspace*{-1.7in}
\caption{Diffuse neutrino flux per flavor (solid black) and $\gamma$-rays (dash-dot) from HNe and SNe, 
for a diffusion coefficient $D\propto \vareps_p^{1/2}$, in both the host galaxy and cluster. 
The proton spectral index is $\Gamma = 2$. The black line with white circles denotes the measured 
atmospheric neutrino flux. The starburst galaxy  contributions are in red lines,
while the normal star-forming galaxy contributions are represented by blue lines.  
The contribution from HNe are in solid lines colored while those from the SNe are dashed. 
The solid green line denotes the total cluster 
contribution (i.e. HNe and SNe from both types of galaxies). Green data points correspond to the 
{Fermi} measurements of the extragalactic diffuse $\gamma$-ray background \citep{Ackermann+15igbfermi}.
Black points correspond to the IceCube measurements of astrophysical neutrinos \citep{IC3+15tevnu}, 
note that two of the low energy data points are within the gray lines of the error bars of the 
atmospheric flux. From \cite{Senno+15clugalnu}}
\end{minipage}
\label{fig:senno15-f1}
\end{figure}
On the other hand, a possible UHECR and TeV-PeV neutrino source closely related to GRBs are 
hypernovae, or jet-driven supernovae  \citep{Wang+07crhn,Budnik+08hn}, and these are in fact 
largely concentrated in star-forming or star-bursting galaxies. While classical long GRBs are often 
associated with hypernovae, there are many more (90-95\%) of hypernovae which do not show signs of 
a GRB, i.e. non-GRB hypernovae are much numerous than GRBs. Hypernovae have also been investigated 
as possible sources of the IceCube diffuse astrophysical TeV-PeV neutrino flux \citep{He+13pevnuhn,
Murase+13pev,Liu+14pevnuhn}. The challenge, with these various sources, is to fit the IceCube
diffuse spectral flux, arising from $pp$ or $p\gamma$ collisions leading to $\pi^+$, while at the 
same time not violating the observed Fermi isotropic diffuse  $\gamma$-ray flux 
\citep{Abdo+10igbfermi,Ackermann+15igbfermi}. These constraints can be reasonably  met  when taking
into account both the hypernova and normal supernova populations in the observed relative ratios,
as seen in Fig. 12, \citep{Senno+15clugalnu}.
The protons accelerated in the hypernovae (HNe) and supernovae (SNe) propagate in the host galaxy
and in the host cluster with a diffusion coefficient $D(\vareps)\propto \vareps^{1/3}$ or 
$\propto \vareps^{1/2}$, where $\vareps$ is proton energy, and make neutrinos and gamma-rays mainly 
by $pp$ collisions, with negligible  $p\gamma$ contributions.  As can be seen in Fig. 12, 
the IceCube flux may be satisfied (except for one point a few sigma away)
if the entire Fermi flux is explained this way. The problem is that AGNs already appears to be
responsible for a large fraction of the diffuse Fermi $\gamma$-ray flux \citep{Abazajian+11igbfermi},
hence not all the IceCube flux may be ascribed to HNe and SNe. Future work needs to clarify the
roles of the various possible sources.
The new all-sky $\gamma$-ray survey instrument HAWC \citep{Westerhoff14hawc} will undoubtedly 
play a large role in the investigation of GRBs \citep{Abeysekara+12hawc,Taboada+14hawcgrb,
Abeysekara+15hawcgrb130427} and other extragalactic as well as galactic sources of cosmic rays 
and neutrinos.

\begin{acknowledgement}
We thank NASA NNX13AH50G for support, and P. Veres and K. Kashiyama for collaborations.
\end{acknowledgement}


\input{msaph.bbl}
\end{document}

%% file: macos.tex
\def\mathnew{\mathsurround=0pt}
\def\simov#1#2{\lower .5pt\vbox{\baselineskip0pt \lineskip-.5pt
       \ialign{$\mathnew#1\hfil##\hfil$\crcr#2\crcr\sim\crcr}}}
\def\simg{\mathrel{\mathpalette\simov >}}
\def\siml{\mathrel{\mathpalette\simov <}}

\def\msun{M_\odot}

\def\fermi{\textit{Fermi}\,}

\def\gbm{\textit{GBM}\,}
\def\lat{\textit{LAT}\,}

\def\vareps{\varepsilon}

\def\beq{\begin{equation}}
\def\enq{\end{equation}}
\def\bea{\begin{eqnarray}}
\def\ena{\end{eqnarray}}
\def\bec{\begin{center}}
\def\enc{\end{center}}

\def\blist{\begin{list}{$\bullet$}{\itemsep 0.0in \parsep 0.0in}}
\def\elist{\end{list}}
\def\bitem{\begin{list}{\arabic{enumi}.}{\usecounter{enumi} \itemsep 0.0in \parsep 0.0in}}
\def\eitem{\end{list}}
\def\cm{\hbox{~cm}}

\def\erg{\hbox{~erg}}

\def\TeV{\hbox{~TeV}}

\def\MeV{\hbox{~MeV}}

\def\Hz{\hbox{~Hz}}

\def\part{\partial}

\def\m-pl{m_{Pl}}

\def\h75{h_{75}}
\def\Omh75{\Omega h^2_{75}}
\def\Omh70{\Omega h^2_{70}}

\def\fun#1#2{\lower3.6pt\vbox{\baselineskip0pt\lineskip.9pt
  \ialign{$\mathsurround=0pt#1\hfil##\hfil$\crcr#2\crcr\sim\crcr}}}

\def\jcap{Jour. Cosmology and Astro-Particle Phys.\,}
\def\mnras{M.N.R.A.S.\,}

\def\apj{Astrophys.J.\,}
\def\apjl{Astrophys.J.Lett.\,}

\def\nat{Nature\,}

\def\prd{Phys.Rev.D\,}
\def\araa{Annu.Rev.Astron.Astrophys.\,}
\def\aap{Astron.Astrophys.\,}
\def\aaps{Astron.Astrophys.Supp.\,}

\def\physrep{Phys.Rep.\,}